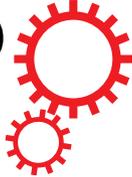

# OPEN  Casimir switch: steering optical transparency with vacuum forces

Xi-fang Liu[1,2], Yong Li[1,3] & H. Jing[1,2]

The Casimir force, originating from vacuum zero-point energy, is one of the most intriguing purely quantum effects. It has attracted renewed interests in current field of nanomechanics, due to the rapid size decrease of on-chip devices. Here we study the optomechanically-induced transparency (OMIT) with a tunable Casimir force. We find that the optical output rate can be significantly altered by the vacuum force, even terminated and then restored, indicating a highly-controlled optical switch. Our result addresses the possibility of designing exotic optical nano-devices by harnessing the power of vacuum.

Cavity optomechanics[1,2], which explores the interaction between electromagnetic waves and mechanical motion, has witnessed rapid advances in recent years, leading to a variety of applications[3], such as high-bandwidth accelerometer[4,5], quantum-limited displacement sensing[6], optical self-focusing[7], quantum transducer[8], and most recently, achieving quantum squeezing of mechanical motion[9]. Another notable example, closely related to the present study, is the experimental demonstration of optomechanically-induced transparency (OMIT)[10–12], which provides a new approach for coherent control of light with a solid device, such as delay or advance of light[13,14], quantum memory[15–17], and precision measurement of tiny objects[18]. The basic mechanism of OMIT is the destructive interference of two absorption channels of the signal photons (i.e. absorbed by the cavity field or the mechanical mode), thereby leading to a transparency window for the signal light in the otherwise strongly absorbed region. This is formally equivalent to that of electromagnetically-induced transparency (EIT) well-known in atomic physics[19]. Further interesting studies on the OMIT include, e.g., nonlinear OMIT[20–23], two-color OMIT[24], cascaded OMIT[25], and reversed OMIT in parity-time resonators[26].

On the other hand, with the unprecedented ability of fabricating and characterizing materials on the nanometer scale, research in exploring and harnessing the exotic quantum effect like the Casimir force (CF)[27,28] has become active in recent years. The CF can become increasingly important in nano-devices as the space separation between the component surfaces is drastically decreased. The high-precision CF measurement was first performed in 1997 by Lamoreaux[29], and then also by several other groups[30–32]. The CF-induced novel effects have been revealed, such as vacuum friction of motion[33–35], non-touching bound of nano-particles[36], nonlinear mechanical oscillations[37], and giant vacuum force near a transmission line[38]. Practical applications of the CF in e.g. quantum sensing of motion was also presented[39], highlighting its impacts on future quantum technologies.

In the present work, by combining these two research fields, we study the CF effect in a cavity optomechanical system. We note that in very recent works, the interplay of the external and zero-point radiation pressures was already investigated for the dynamics of a levitated nanosphere trapped in a cavity[40,41]. Here we focus on the OMIT process in the presence of a tunable vacuum force. We find an interesting CF-controlled optical switch effect, i.e. the optical transparency window can be completely shut down and also re-opened again by sorely tuning the strength of the Casimir force. To be more specific, the presence of the CF leads to the modifications of not only the steady-state values of the dynamical variables, but also the field fluctuations and subsequently, the optical output rate of the probe light. In particular, for a fixed sphere spatially separated from the moveable mirror, by reducing the air-gap distance, the conventional OMIT spectrum tends to be shifted to the red-detuning side (with some distortions as well). As a result, by tuning the CF, the output of the probe light at the probe-cavity resonance can be attenuated, or even totally shut down and then restarted again, for a fixed pump power. In addition, we find that even for the non-OMIT case, i.e. without any pump light, the CF-aided optical transparency can still be achieved in our proposed situation. A reversed pump-dependence was also revealed for the CF-aided OMIT, for the low-power cases, in comparison with the conventional OMIT. These results indicate the possibility of designing unconventional optical nano-devices by exploiting vacuum zero-point energy.

[1]Key Laboratory of Low-Dimensional Quantum Structures and Quantum Control of Ministry of Education, Department of Physics and Synergetic Innovation Center for Quantum Effects and Applications, Hunan Normal University, Changsha 410081, China. [2]Department of Physics, Henan Normal University, Xinxiang 453007, China. [3]Beijing Computational Science Research Center, Beijing 100084, China. Correspondence and requests for materials should be addressed to H.J. (email: jinghui73@gmail.com)

## Results

### The model of the system.
We consider a cavity optomechanical system with a tunable CF. The optical cavity mode, characterized by the resonance frequency $\omega_c$ and the decay rate $\gamma$, is coupled to the moveable mirror via the optomechanical coupling rate $g = \omega_c/L$ ($L$ is the cavity length); also the moveable mirror interacts with a nearby gold-coated nanosphere via the CF. For a fixed sphere-plate separation $d$ of perfect conductors, the zero-point CF is given by[42,43]

$$F_C^{(T=0)} = \frac{2\pi^3 \hbar c R}{720 d^3}\left[1 + \frac{d}{2R}(2\beta - 1)\right], \quad \left(\beta = \frac{2}{3} - \frac{10}{\pi^2} \sim -0.35\right). \quad (1)$$

where the first term is the perfect reflector formula in the proximity force approximation (PFA), the second term accounts for the leading correction to the PFA[43], and $c$ is the light speed at vacuum, $R$ is the radius of the sphere. The condition $d/R \ll 1$ is the standard condition that determines the validity of the PFA; for $d/R \ll 1$, the second term is safely neglected. The thermal CF, $F_C^{(T)}$, dominates for large separations $d \gtrsim 3\,\mu m$, but is much smaller than $F_C^{(T=0)}$ for $d \gtrsim 1\,\mu m$[42]. Here we focus on the latter regime $F_C^{(T)} \ll F_C^{(T=0)}$. In current experiments, the CF has been accurately measured for $d \sim 100$ nm, still showing excellent agreement with theoretical predictions[44,45]. Complicated calculations of various non-ideal CF corrections have also been developed[42,43], leading to e.g. an increase of about 1% in the CF due to the surface roughness, for a torsion balance experiment[42]; for numerical calculations of the CF with finite conductivity, confirming the validity of the plasma model for the gold, see ref. 27. We stress that in this work we focus on the vaccum-assisted steering of OMIT spectrum, instead of various non-ideal CF corrections (for these efforts, see e.g. ref. 27).

The cavity is driven by a strong control laser with the frequency $\omega_L$ and a weak probe laser with the frequency $\omega_p$. The field amplitudes of these two lasers are given by, respectively,

$$\varepsilon_L = \sqrt{\frac{2 P_L \gamma}{\hbar \omega_L}}, \quad \varepsilon_p = \sqrt{\frac{2 P_{in} \gamma}{\hbar \omega_p}},$$

where $P_L$ and $P_{in}$ denote the powers of the pump and the probe lasers. In the frame rotating at the frequency $\omega_L$, the Hamiltonian of this CF-aided optomechanical system can be written at the simplest level as[1–3,12]

$$\begin{aligned}
H &= H_0 + H_{int} + H_C + H_{dr}, \\
H_0 &= \hbar \Delta_L a^\dagger a + \left(\frac{p_1^2}{2m_1} + \frac{1}{2}m_1 \omega_{m,1}^2 x_1^2\right) + \left(\frac{p_2^2}{2m_2} + \frac{1}{2}m_2 \omega_{m,2}^2 x_2^2\right), \\
H_{int} &= -\hbar g a^\dagger a x_1, \\
H_C &= -\frac{\pi^3 \hbar c R}{720[d - (x_1 - x_2)]^2}, \\
H_{dr} &= i\hbar[\varepsilon_L(a^\dagger - a) + \varepsilon_p(a^\dagger e^{-i\nu t} - a e^{i\nu t})].
\end{aligned} \quad (2)$$

Here $a$ and $a^\dagger$ are the creation and annihilation operators of the cavity mode respectively, $m_i$ or $\omega_{m,i}$ ($i = 1, 2$) denotes the mass or resonance frequency of the oscillator respectively, the optical detuning terms are

$$\Delta_L = \omega_c - \omega_L, \quad \Delta_p = \omega_p - \omega_c, \quad \nu = \omega_p - \omega_L,$$

and $p_i$ or $x_i = x_{0,i}(b_i + b_i^\dagger)$ denotes the momentum or position operator of the mechanical oscillator respectively, with $x_{0,i} = \sqrt{\frac{\hbar}{2 m_i \omega_{m,i}}}$ and the phonon-mode operators $b_i, b_i^\dagger$ (see Fig. 1). We first focus on the single oscillator case, and then discuss the case with coupled two oscillators.

### The fixed-sphere case.
For $x_1 \to x$, $x_2 \to 0$, the Heisenberg equations of motion are

$$\dot{a} = (-i\Delta_L + igx - \gamma)a + \varepsilon_L + \varepsilon_p e^{-i\nu t}, \quad (3)$$

$$\ddot{x} + \Gamma_m \dot{x} + \omega_m^2 x + \frac{2 V_{sp}}{m(d - x)^3} = \frac{\hbar g}{m} a^\dagger a, \quad (4)$$

where $V_{sp} = -\frac{\pi^3 \hbar c R}{720}$, $\Gamma_m$ is the mechanical decaying rate of the vibrating mirror, and we have neglected the noise terms and taken $m_1 \to m$, $\omega_{m,1} \to \omega_m$, for simplicity. The steady-state values of the dynamical variables are

$$a_s = \frac{\varepsilon_L}{\gamma + i(\Delta_L - g x_s)}, \quad x_s = \frac{1}{m \omega_m^2}\left[\hbar g |a_s|^2 - \frac{2 V_{sp}}{(d - x_s)^3}\right].$$

For experimentally-accessible values of the system parameters[46,47], i.e. $\omega_c = 2\pi c/\lambda$, $\lambda = 1064$ nm, $L = 25$ mm, $R = 150$ nm, $m = 145$ ng, $\gamma = 2\pi \times 80$ KHz, $\Gamma_m = 2\pi \times 141$ Hz, and $\Delta_L = \omega_m = 2\pi \times 947$ KHz, we find $d/R \ll 1$ (for $d < 5\,\text{nm}$) and $x_s/d \lesssim 10^{-2}$ (for $d > 1.5\,\text{nm}$). Hence it is good enough to take the Casimir term up to the second order of $x_s/d$, within our interested regime, as numerically confirmed later. Under this approximation, we have the balance equation of the moveable mirror

$$m\omega_m^2 x_s + \left(\frac{6V_{sp}}{d^4}x_s + \frac{2V_{sp}}{d^3}\right) = \hbar g |a_s|^2, \quad (5)$$

where the first (second) term in the left-hand side results from the restoring (Casimir) force. Clearly, the left-hand side should be positive, which is fulfilled for $d > 0.7\,\text{nm}$, with the above parameters. Less values of $d$ lead to a CF stronger than the restoring force, and thus adhesion of the mirror. We also find that the CF term is much weaker than the restoring force for $d > 10\,\text{nm}$; in contrast, it becomes comparable with the latter for $0.7\,\text{nm} < d \lesssim 2\,\text{nm}$. We note that in current experiments, the CF measurements for $d = 2\,\text{nm}$ are challenging; nevertheless, even for a larger $d$, it is still possible to achieve the required strong CF by altering optical properties or geometric structures of the interacting materials[31,38,48–54]. For examples, for parallel graphene layers with $d < 10\,\text{nm}$, it was found that the CF $\sim d^{-5}$; with specific nonostructures, further enhancement as CF $\sim d^{-7}$ can be achieved[54]. This indicates that the required CF, corresponding to $d \sim 1\,\text{nm}$ for ideal metals, can be achieved for larger values of $d$, e.g. $d \sim 10\,\text{nm}$ or even $50\,\text{nm}$, by proper designs of material properties. In fact, there is a huge list of materials whose electromagnetic response can be widely tuned, hence allowing for significant CF enhancement at fixed separations, e.g. optical crystals, semiconductors, topological insulators, or plasmonic nanostructures (see ref. 54 for a very recent review).

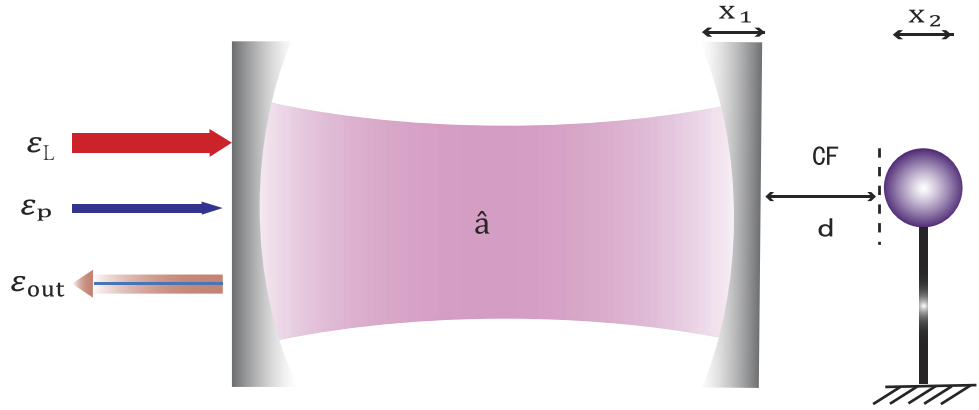

**Figure 1. Optomechanical system with a tunable Casimir force (CF).** The Fabry-Pérot cavity contains a moveable mirror, which interacts with both the cavity field and the nearby gold-coated nanosphere via radiation pressures. The external surface of the mirror is also gold-coated. The nanosphere is either fixed or moveable (e.g. by attaching it to a cantilever).

Here we show that a novel CF-controlled optical switch can be achieved in an OMIT system, even in the low-power linear regime. In order to see this, we expand each operator as the sum of its steady-state value and a small fluctuation around that value, i.e., $a = a_s + \delta_a$, $x = x_s + \delta_x$. After eliminating the steady-state values, we obtain the linearized equations

$$\delta\dot{a} = (-i\Delta_L + igx_s - \gamma)\delta a + iga_s\delta x + \varepsilon_p e^{-i\nu t}, \quad (6)$$

$$\delta\ddot{x} + \Gamma_m \delta\dot{x} + \left[\omega_m^2 + \frac{6V_{sp}}{m(d-x_s)^4}\right]\delta x = \frac{\hbar g}{m}(a_s^* \delta a + a_s \delta a^\dagger). \quad (7)$$

By applying the ansatz[10–12,14,26]

$$\begin{pmatrix} \delta a \\ \delta x \end{pmatrix} = \begin{pmatrix} \delta a_+ \\ X \end{pmatrix} e^{-i\nu t} + \begin{pmatrix} \delta a_- \\ X^* \end{pmatrix} e^{i\nu t}, \quad (8)$$

the linearized equations are transformed into

$$(i\Delta_L - igx_s + \gamma - i\nu)\delta a_+ = iga_s X + \varepsilon_p, \quad (9)$$

$$(i\Delta_L - igx_s + \gamma + i\nu)\delta a_- = iga_s X^*, \quad (10)$$

$$(\Omega_m^2 - \nu^2 - i\nu\Gamma_m)X = \frac{\hbar g}{m}(a_s^* \delta a_+ + a_s \delta a_-^\dagger), \quad (11)$$

with an effective mechanical frequency

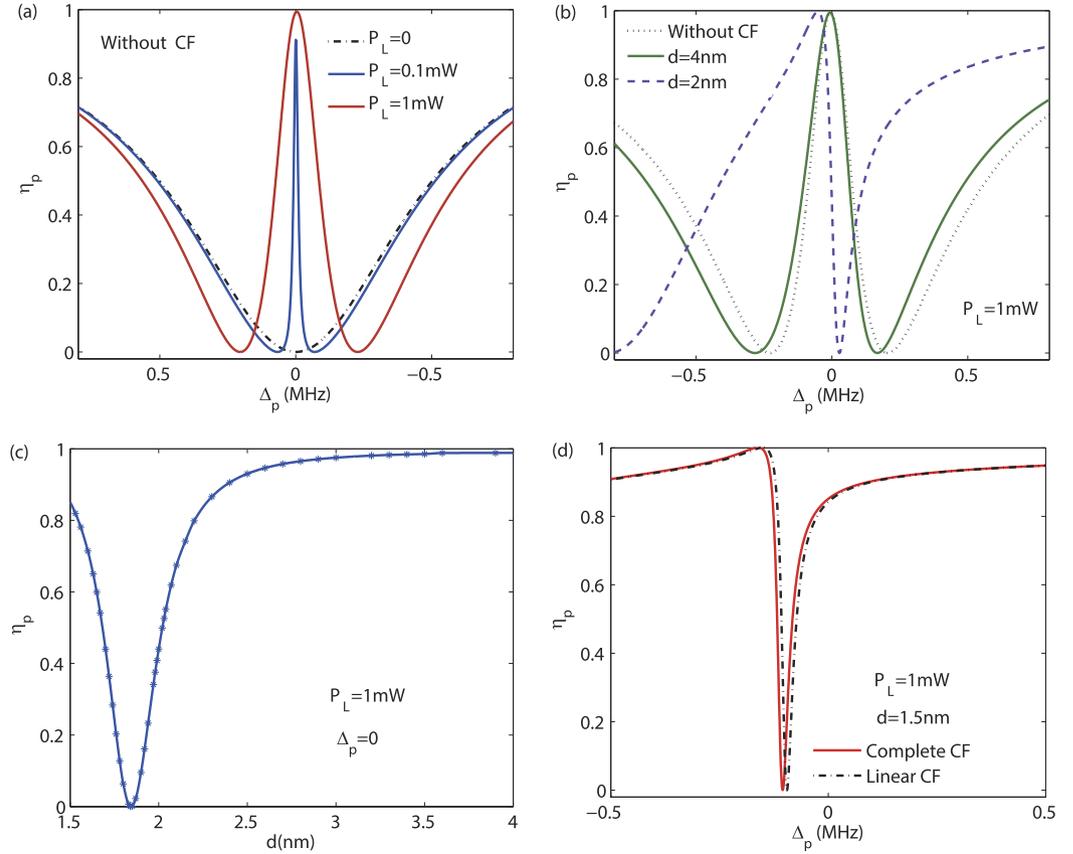

**Figure 2.** (**a,b,d**) The output rate $\eta_p$ of the probe light versus the optical detuning $\Delta_p = \omega_p - \omega_c$, for different values of the mirror-sphere separation $d$. (**c**) shows the dependence of $\eta_p$ (at the resonance $\Delta_p = 0$ on $d$, indicating a CF-controlled light switch. See the text for the values of other parameters.

$$\Omega_m = \sqrt{\omega_m^2 + \frac{6V_{sp}}{m(d-x_s)^4}}.$$

These equations can then be solved as

$$\delta a_+ = \frac{[(\Omega_m^2 - \nu^2 - i\nu\Gamma_m)G_+ m + i\hbar g^2 n_s]\varepsilon_p}{(\Omega_m^2 - \nu^2 - i\nu\Gamma_m)G_+ G_- m - i\hbar g^2 (G_+ - G_-)n_s}, \tag{12}$$

$$\delta a_- = \frac{i\hbar g^2 n_s \varepsilon_p}{(\Omega_m^2 - \nu^2 + i\nu\Gamma_m)G_+^* G_-^* m + i\hbar g^2 (G_+^* - G_-^*)n_s}, \tag{13}$$

$$X = \frac{\hbar g a_s^* G_+ \varepsilon_p}{(\Omega_m^2 - \nu^2 - i\nu\Gamma_m)G_+ G_- m - i\hbar g^2 (G_+ - G_-)n_s}, \tag{14}$$

where $G_\pm = \gamma \mp i\Delta_L \pm igx_s - i\nu$, and $n_s = |a_s|^2$ is the intracavity photon number. The expectation value of the output optical field can be obtained by using the standard input-output relation, i.e. $a^{out}(t) = a^{in} - \sqrt{2\gamma}a(t)$, where $a^{in}(t)$ and $a^{out}(t)$ are the input and output field operators. This leads to the optical reflection rate for the probe field, i.e. the amplitude square of the ratio of the output field amplitude to the input field amplitude at the probe field frequency,

$$\eta(\omega_p) = \left|1 - \left(\frac{2\gamma}{\varepsilon_p}\right)\delta a_+\right|^2. \tag{15}$$

We calculate this rate to better understand the CF-aided OMIT process under the above parameters as well as $\Delta_L = \omega_m$ and $\Delta_p \equiv \omega_p - \omega_c = \nu - \omega_m$. As Fig. 2 shows, for the conventional OMIT without any CF, at the res-onance $\Delta_p = 0$, the probe light is absorbed for $P_L = 0$, while it becomes transparent by applying a pump light [see Fig. 2(a)]; in contrast, for the CF-aided OMIT, we find that by reducing $d$, $\eta_p(\Delta_p = 0)$ is firstly decreased until zero (at $d \sim 1.8$ nm) and then increased again [see Fig. 2(b,c), for a fixed value of $P_L = 1$ mW]. This indicates a CF-controlled light switch, even shutting down and re-starting the signal [see Fig. 2(c)]. One mechanism underlying this effect is the following: the CF-induced frequency shift $\omega_m \to \Omega_m$ modifies the resonance condition as $\widetilde{\Delta}_p \equiv \nu - \Omega_m = 0$, or correspondingly, $\Delta_p = \nu - \omega_m = \Omega_m - \omega_m < 0$, i.e. the OMIT spectrum tends to be shifted to the left. This effect is also reminiscent of that using the electrostatic force to tune the OMIT[55] or that with an external mechanical driving[25]. Also Fig. 2(d) shows the result about $\eta_p$ with a linearized CF, indicating that the CF-controlled light switch works well even in the linear CF regime.

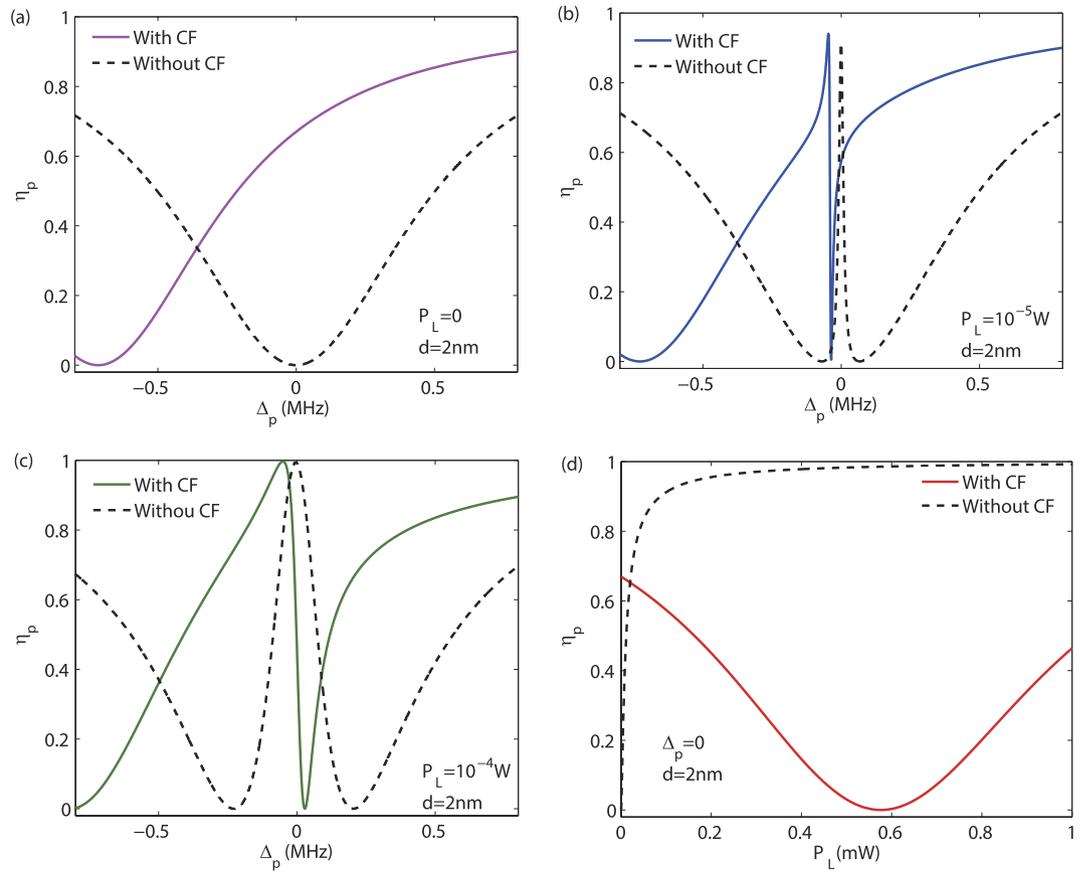

**Figure 3.** (**a**) The CF-assisted OMIT, without any pump field; (**b,c**) the output rate $\eta_p$ versus the detuning $\Delta_p$, for weak values of $P_L$ (see also ref. 56); (**d**) reversed pump dependence of $\eta_p$, at the resonance $\Delta_p = 0$, in the low-power CF-aided OMIT. For all the cases with the CF, we take a fixed value of $d = 2$ nm as a typical example. All the other parameter values are the same as in Fig. 2.

Interestingly, we find that even for $P_L = 0$, the probe light can become transparent by steering the CF [at $\Delta_p = 0$, see Fig. 3(a)]. In this situation, the vacuum field, instead of the pump field, serves as the control gate for the output of the probe light. For weak pump powers, the CF-aided OMIT shows an exotic feature of reversed pump dependence at $\Delta_p = 0$, in comparison with the conventional OMIT [see Fig. 3(b–d)] and also ref. 56. These results show that (i) even without any pump field, the signal light can still be transparent with the aid of the virtual photons (i.e. the Casimir potential); (ii) combining the real-photon (e.g. the pump light) and the virtual-photon (i.e. the vacuum fluctuation) fields provide more flexible and efficient ways to manipulate the light propagation.

**The moveable-sphere case.** For completeness, we also consider a nanosphere attached to a vibrating cantilever. We note that this configuration was recently exploited to design a Casimir parametric amplifier[56]. Since the linearized CF was already confirmed to be a good approximation in our system, we can expand $H_C$ in Eq. (1) up to the quadratic term of $(x_1 - x_2)^2$. The linear term $x_{1,2}$ and the quadratic term $x_{1,2}^2$ can be absorbed into the re-defined equilibrium positions and the mechanical frequency, respectively, and thus are unimportant; the term of interests is the inter-mode coupling

$$H_C = \hbar J x_1 x_2, \tag{16}$$

with $J = \frac{\pi^3 cR}{120 d^4}$. This kind of coupling has been achieved in various physical systems, to facilitate e.g. quantum state transfer of two spatially separated oscillators[57–60].

The equations of motion of the resulting three-mode system are

$$\dot{a} = (-i\Delta_L + igx_1 - \gamma)a + \varepsilon_L + \varepsilon_p e^{-i\nu t}, \quad (17)$$

$$\ddot{x}_1 + \Gamma_{m,1}\dot{x}_1 + \omega_{m,1}^2 x_1 = \frac{\hbar g}{m_1}a^\dagger a - \frac{\hbar J x_2}{m_1}, \quad (18)$$

$$\ddot{x}_2 + \Gamma_{m,2}\dot{x}_2 + \omega_{m,2}^2 x_2 = -\frac{\hbar J x_1}{m_2}, \quad (19)$$

from which we obtain the steady-state values of the dynamical variables

$$a_s = \frac{\varepsilon_L}{\gamma + i(\Delta_L - gx_{1,s})}, \quad x_{1,s} = \frac{\hbar(g|a_s|^2 - Jx_{2,s})}{m_1\omega_{m,1}^2}, \quad x_{2,s} = -\frac{\hbar J x_{1,s}}{m_2\omega_{m,2}^2}. \quad (20)$$

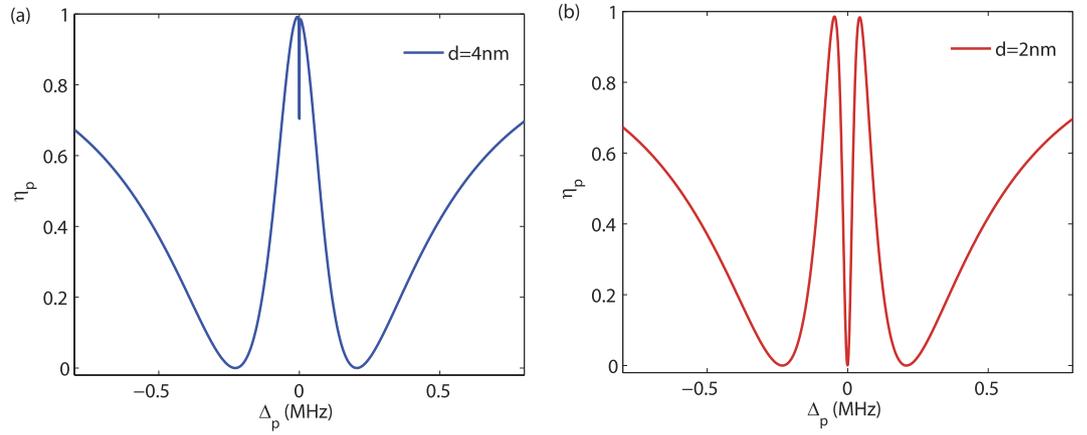

**Figure 4. The output rate $\eta$ of the probe light versus the optical detuning $\Delta_p$, with different values of $d$.** Here we take $P_L = 1$ mW, and assume the same parameters for the two mechanical oscillators, just for simplicity. All the other parameters are the same as in Fig. 2.

Then by following the procedure as above, we have the linearized equations of motion and their solutions. The final result about $\eta_p$ is plotted in Fig. 4.

For $d \to \infty$, as Fig. 2(a) shows, we have the conventional OMIT spectrum, i.e. a single-peak transparency at $\Delta_p = 0$. In contrast, for the CF-aided OMIT with two coupled mechanical oscillators, a dip emerges at $\Delta_p = 0$ [for $d = 4$ nm, see Fig. 4(a)] and the OMIT window is then split into a double-peak structure [for $d = 2$ nm, see Fig. 4(b)]. Hence, by tuning the CF, the probe light can be varied from the transparency regime to the absorption regime, or vice versa. The shape of the OMIT spectrum as Fig. 4 is very similar to that in an Autler-Townes splitting (ATS) situation[61,62], and the relation between these two kinds of physical processes, or even the controllable transition of them, would be an interesting problem to be explored in our future works. We also remark that the coupling as in Eq. (16) can also be realized by using e.g. coupled charged objects, with which a similar OMIT spectrum was observed[55]. As a comparison, our proposal here focuses on the single-oscillator case, instead of the double-oscillator case as studied in ref. 55; more importantly, it does not require any charged or magnetic object. The Casimir force comes from the vacuum itself and plays a crucial role in chip-scale nano-devices with decreasing vacuum distances between different elements[28,39,45].

## Methods
**Derivation of the optical output rate for the moveable sphere.** Taking the expectation of each operator given in Eqs.(17)-(19), we find the linearized Heisenbrg equations as

$$\delta\dot{a} = (-i\Delta_L + igx_{1,s} - \gamma)\delta a + iga_s \delta x_1 + \varepsilon_p e^{-i\nu t}, \quad (21)$$

$$\delta\ddot{x}_1 + \Gamma_{m,1}\delta\dot{x}_1 + \omega_{m,1}^2 \delta x_1 = \frac{\hbar g}{m_1}(a_s^* \delta a + a_s \delta a^\dagger) - \frac{\hbar J}{m_1}\delta x_2, \quad (22)$$

$$\delta\ddot{x}_2 + \Gamma_{m,2}\delta\dot{x}_2 + \omega_{m,2}^2 \delta x_2 = -\frac{\hbar J}{m_2}\delta x_1. \quad (23)$$

by applying the following ansatz,

$$\delta x_1 = X_1 e^{-i\nu t} + X_1^* e^{i\nu t}, \quad (24)$$

$$\delta x_2 = X_2 e^{-i\nu t} + X_2^* e^{i\nu t}, \tag{25}$$

$$\delta a = \delta a_+ e^{-i\nu t} + \delta a_- e^{i\nu t}. \tag{26}$$

equations (21)–(23) can be transformed into the following form,

$$(i\Delta_L - igx_{1,s} + \gamma - i\nu)\delta a_+ = iga_s X_1 + \varepsilon_p, \tag{27}$$

$$(i\Delta_L - igx_{1,s} + \gamma + i\nu)\delta a_- = iga_s X_1^*, \tag{28}$$

$$(\omega_{m,1}^2 - \nu^2 - i\nu\Gamma_{m,1})X_1 = \frac{\hbar g}{m_1}(a_s^* \delta a_+ + a_s \delta a_-^\dagger) - \frac{\hbar J}{m_1}X_2, \tag{29}$$

$$(\omega_{m,1}^2 - \nu^2 + i\nu\Gamma_{m,1})X_1^* = \frac{\hbar g}{m_1}(a_s \delta a_+^\dagger + a_s^* \delta a_-) - \frac{\hbar J}{m_1}X_2^*, \tag{30}$$

$$(\omega_{m,2}^2 - \nu^2 - i\nu\Gamma_{m,2})X_2 = -\frac{\hbar J}{m_2}X_1, \tag{31}$$

$$(\omega_{m,2}^2 - \nu^2 + i\nu\Gamma_{m,2})X_2^* = -\frac{\hbar J}{m_2}X_1^*. \tag{32}$$

Solving these algebraic equations leads to

$$\delta a_+ = \frac{(a_1 a_2 m_1 m_2 G_2 + i\hbar g^2 a_2 m_2 n_s - \hbar^2 J^2 G_2)\varepsilon_p}{a_1 a_2 m_1 m_2 G_1 G_2 - i\hbar g^2 a_2 m_2 n_s (G_2 - G_1) - \hbar^2 J^2 G_1 G_2}, \tag{33}$$

$$\delta a_- = \frac{i\hbar g^2 a_2 m_2 a_s^2 \varepsilon_p}{a_1^* a_2^* m_1 m_2 G_1^* G_2^* + i\hbar g^2 a_2^* m_2 n_s (G_2^* - G_1^*) - \hbar^2 J^2 G_1^* G_2^*}, \tag{34}$$

$$X_1 = \frac{\hbar g a_2 m_2 a_s^* G_2 \varepsilon_p}{a_1 a_2 m_1 m_2 G_1 G_2 - i\hbar g^2 a_2 m_2 n_s (G_2 - G_1) - \hbar^2 J^2 G_1 G_2}, \tag{35}$$

$$X_2 = \frac{-\hbar^2 gJ a_s^* G_2 \varepsilon_p}{a_1 a_2 m_1 m_2 G_1 G_2 - i\hbar g^2 a_2 m_2 n_s (G_2 - G_1) - \hbar^2 J^2 G_1 G_2}. \tag{36}$$

where we have used $n_s = |a_s|^2$ and

$$a_1 = \omega_{m,1}^2 - \nu^2 - i\nu\Gamma_{m,1}, \tag{37}$$

$$a_2 = \omega_{m,2}^2 - \nu^2 - i\nu\Gamma_{m,2}, \tag{38}$$

$$G_1 = i\Delta_L - igx_{1,s} + \gamma - i\nu, \tag{39}$$

$$G_2 = -i\Delta_L + igx_{1,s} + \gamma - i\nu. \tag{40}$$

the expectation value $\langle a^{out}(t) \rangle$ of the output field $a^{out}(t)$ can be calculated using the standard input-output relation $a^{out}(t) = a^{in} - \sqrt{2\gamma}\, a(t)$, where $a^{in}(t)$ and $a^{out}(t)$ are the input and output field operators, and

$$\langle a^{out}(t) \rangle = \left[\frac{\varepsilon_L}{\sqrt{2\gamma}} - \sqrt{2\gamma}(a_s + \delta a_-)e^{i\nu t}\right]e^{-i\omega_L t} + \left(\frac{\varepsilon_p}{\sqrt{2\gamma}} - \sqrt{2\gamma}\delta a_+\right)e^{-i(\omega_L+\nu)t}. \tag{41}$$

Hence, the output rate of the probe field can be written as $\eta(\omega_p) = |t(\omega_p)|^2$, where $t(\omega_p)$ is the ratio of the output field amplitude to the input field amplitude at the probe frequency.

$$t(\omega_p) = \frac{\varepsilon_p - 2\gamma\delta a_+}{\varepsilon_p}. \tag{42}$$

## Conclusion

In summary, we have demonstrated the effects of the vacuum force on the OMIT, indicating the possibility of controlling light with the vacuum. We note that the measurements of the CF for a very short distance are still missing in current experiment; however, even for a constant distance, it is still possible to significantly enhance the CF by, e.g. calibrating an unconventional surface structure[38] or engineering optical properties of novel materials[49–54,63]. With rapid advances of nano-calibration techniques and very active efforts on controlling or enhancing the CFs, our proposal holds the promise to be realized, at least in principle. In comparison with a recent work on tuning the OMIT with a voltage-controlled electrostatic force[18], our proposal here does not need any charged or magnetic object, since the CF comes from the *vacuum* itself, which can be of increasingly important in chip-scale nano-devices with decreasing vacuum spaces between the elements. We also note that recently in an optomechanical system, a new kind of motion-induced few percentage correction to the CF was revealed[64,65], indicating that more interesting works could be performed by combining optomechanics and the CF. In the future, we plan to also study the CF-controlled slow light, the cascaded OMIT with coupled Casimir oscillators[25,45], and the CF-mediated quantum mechanical squeezing.

### Acknowledgements

We thank Chang-Pu Sun, Franco Nori, and Yu-xi Liu for discussions. This work is supported partially by the CAS 100-Talent program and the National Natural Science Foundation of China under Grand numbers 11274098, 11474087, and 11422437.

### Author Contributions

H.J. conceived the idea and performed the calculations with the aid of X.F.L. and H.J. wrote the manuscript with the input of Y.L.; and all the authors discussed the content of the manuscript.

### Additional Information

**Competing financial interests:** The authors declare no competing financial interests.

**How to cite this article**: Liu, X.-F. *et al.* Casimir switch: steering optical transparency with vacuum forces. *Sci. Rep.* **6**, 27102; doi: 10.1038/srep27102 (2016).